\documentstyle[12pt]{article}
\newcommand{\beq}{\begin{equation}}
\newcommand{\eeq}{\end{equation}}
\newcommand{\eq}[1]{(\ref{#1})}
\newcommand{\A}{{\bf a}}
\newcommand{\B}{{\bf b}}
\newcommand{\C}{{\bf c}}
\newcommand{\beqn}{\begin{eqnarray}}
\newcommand{\eeqn}{\end{eqnarray}}
\newcommand{\dst}{&\displaystyle}
\newcommand{\al}{\mbox{$Z\alpha$}}
\newcommand{\eps}{\mbox{$\varepsilon$}}
\newcommand{\Q}{\mbox{$\kappa$}}
\newcommand{\r}{\mbox{${\bf r}$}}
\newcommand{\k}{\mbox{${\bf k}$}}

\newcommand{\g}{\mbox{\boldmath ${\gamma}$\unboldmath}}

\newcommand{\e}{\mbox{${\bf e}$}}
\newcommand{\bi}[1]{\bibitem{#1}}
\newcommand{\fr}[2]{\frac{#1}{#2}}
\newcommand{\p}{\mbox{${\bf p}$}}

\newcommand{\q}{\mbox{${\bf q}$}}
\newcommand{\vq}{\mbox{${\bf Q}$}}
\newcommand{\vx}{\mbox{${\bf x}$}}
\newcommand{\vd}{\mbox{${\bf \Delta}$}}
\newcommand{\f}{\mbox{${\bf f}$}}
\begin{document}
\begin{titlepage}

\begin{center}
{\Large \bf Budker Institute of Nuclear Physics}
\end{center}

\vspace{1cm}

\begin{flushright}
{\bf Budker INP 98-28\\
April 15, 1998 }
\end{flushright}

\vspace{1.0cm}
\begin{center}{\Large \bf Cross section of high-energy photon
splitting in the electric fields of heavy atoms.}\\
\vspace{1.0cm}
{\bf R.N. Lee, A.I. Milstein, V.M. Strakhovenko} \\
G.I. Budker Institute of Nuclear Physics,\\
630090 Novosibirsk, Russia\\

\vspace{4.0cm}
\end{center}

\begin{abstract}
Various differential cross sections of high-energy photon splitting
in the electric fields of heavy atoms are calculated exactly in the
parameter \al. The consideration is based on the quasiclassical
approach applicable for small angles between all photon momenta. The
expressions obtained are valid for arbitrary transverse momenta of
final photons . The detailed investigation of the process is
performed taking into account the effect of screening .
The exact cross section turns out to be noticeably smaller than the
result obtained in the Born approximation.
\end{abstract}
\end{titlepage}

\newpage

\section{\bf Introduction}

The first successful observation of
high-energy photon splitting in the electric fields of atoms has been recently
performed in the Budker Institute of Nuclear Physics. A crystal of
$\rm Bi_4Ge_3O_{12}$  has been used as a target. At the present time,
the data processing is almost completed and preliminary results
are published in \cite{Budker}. Theoretical and experimental
investigation of this nonlinear QED process is important as a new
test of QED in strong external fields. It also gives a possibility to
understand the role and
the structure of higher orders of the perturbation theory with respect
to the external field since, as is shown in the
present paper, the exact in $\al$ cross section ($Z|e|$ is the nucleus
charge, $\alpha =\, e^2/4\pi\, =1/137$ is the fine-structure
constant, $e$ is the electron charge) differs essentially from that
obtained in the Born approximation, i.e. in the lowest order in \al.

 The cross section of photon splitting
was found in \cite{Sh,CTP} in the Born approximation. In the
same approximation, an essentially
simpler form of the cross section was obtained in \cite{B} with the
help of the Weizs\"acker-Williams method providing the
logarithmic accuracy. Using the analytical results of \cite{Sh,CTP},
the cross section of the photon splitting was investigated
numerically in \cite{JMO,S}.

 The Coulomb corrections represent the difference between
 the exact ( with exact account for an external field ) cross section of
 the process and the result obtained in the Born
 approximation. Basing on our experience concerning Delbr\"uck scattering, we
 expected a measureable effect in the photon splitting too. In recent papers
\cite{LMS1,LMS2} general formulae for the high-energy
photon splitting amplitudes have been derived exactly in $\al$ for
small angles $f_2$ and $f_3$ between the momenta $\k_2$ , $\k_3$ of
the final photons and the momentum $\k_1$ of the initial one (
$\omega_i=|\k_i|\,\gg m$ , $m$ is the electron mass). It is the
region of small angles, which makes the main contribution to the
total cross section of the process. Additionally, small angles and high
photon energies allow one to apply the quasiclassical approach,
developed in \cite{MS,LM} at the consideration of Delbr\"uck
scattering (see recent review \cite{MShu}).  This approach is based
on the use of the quasiclassical  Green function for the Dirac
equation in the electric field of an atom, which is equivalent to a
summation for all orders of the perturbation theory with respect to
the external field. The quasiclassical approach greatly
simplifies the calculation of amplitudes.

First theoretical results concerning Coulomb corrections in the process of
 photon splitting were obtained in\cite{LMS2} for
 large transverse momenta of the final
photons as compared to the electron mass: $|\k_{2\perp}|=\omega_2f_2
\gg m\,$ and
$|\k_{3\perp}|=\omega_3f_3 \gg m\,$. It turned out that in this
kinematical region the Coulomb corrections  lead to the significant
decrease of the cross section. They
 become noticeable starting from relatively small $Z$ and reach several
tens per cent for heavy atoms.

In the present paper we obtain the exact cross section of the process
in a simple form, valid for arbitrary transverse momenta of the final
photons . Using this result, we examine
numerically the role of the Coulomb corrections in various
differential cross sections. We discuss the case of a pure Coulomb
field as well as the effect of screening.

\section{\bf Amplitudes of the process}

As was shown in \cite{LMS1,LMS2}, it is convenient to present the
initial expression for the photon splitting amplitude in the form
containing the Green functions $D(x,x\,')$ of the "squared" Dirac
equation:
$$D(x,x\,')=\langle x|1/(\hat{\cal P}^2-m^2+i0)|x\,'\rangle\ ,$$
where $\hat{\cal P}=\gamma^{\mu}(i\partial_{\mu}- g_{\mu 0}U(r))$,
$U(r)$ is the potential energy of an electron in an external field,
$\gamma^{\mu}$ are the Dirac matrices. Then the amplitude $M$ is
splitted into a sum of the perturbation-theory diagrams, containing
either three or two Green functions $D(x,x\,')$:
$M=M^{(3)}+M^{(2)}$. The term $M^{(3)}$ is given by
\beqn\label{M33}
\dst
M^{(3)}=\fr{i}{2}{e^3}\int\fr{d\eps}{2\pi}\int d\r_1d\r_2d\r_3
\exp[i(\k_1\r_1-\k_2\r_2-\k_3\r_3)]\times\\
\dst
\mbox{Tr}\biggl\{[(-\hat e_1\hat k_1-2\e_1\p)D(\r_1,\r_2\,|\eps-\omega_2)]
[(\hat e_2^*\hat k_2-2\e_2^*\p)D(\r_2,\r_3\,|\eps)]\times\nonumber\\
\dst
[(\hat e_3^*\hat k_3-2\e_3^*\p)D(\r_3,\r_1\,|\eps+\omega_3)]\biggr\}
+\,(k_2^{\mu}\leftrightarrow k_3^{\mu}\, ,\,
\e_2\leftrightarrow\e_3)\, .\nonumber
\eeqn

Here $e_1^{\mu}$ and $e_{2,3}^{\mu}$ are the polarization vectors of
the initial and final photons, $\hat e = e^{\mu}\gamma_{\mu}=-\e\g$,
and the operator $\p=-i\mbox{\boldmath ${\nabla}$\unboldmath}$
differentiates the Green function $D$ with respect to its first
argument. The term $M^{(2)}$ reads
\beqn\label{M32}
\dst
M^{(2)}=i{e^3}\int\fr{d\eps}{2\pi}\int d\r_1d\r_2\, \mbox{Tr}
\biggl\{\exp[i(\k_1\r_1-\k_2\r_2-\k_3\r_2)]\, \e_2^*\e_3^*\,
\times  \\
\dst
[(-\hat e_1\hat k_1-2\e_1\p)D(\r_1,\r_2\,|\eps-\omega_1)]
\,D(\r_2,\r_1\,|\eps)]\, + \biggl[
\exp[i(\k_1\r_1-\k_2\r_2-\k_3\r_1)]\, \e_1\e_3^*\,\times \nonumber\\
\dst
D(\r_1,\r_2\,|\eps-\omega_2)]\,
[(\hat e_2^*\hat k_2-2\e_2^*\p)D(\r_2,\r_1\,|\eps)]\, +
\,(k_2^{\mu}\leftrightarrow k_3^{\mu}\, ,\, \e_2\leftrightarrow\e_3)
\,\biggr]\biggr\}\, .\nonumber
\eeqn

As was pointed out in \cite{LMS1}, the effect of screening  is
important only for the lowest (Born) approximation in \al. Therefore,
we start from  the case of a pure Coulomb
potential $U(r)=-\al/r$ and then multiply the Born contribution
to the amplitude by
the atomic form factor to take the effect of screening into account.

It is convenient to perform the calculations in terms of the
helicity amplitudes\\
$M_{\lambda_1\lambda_2\lambda_3}(\k_1,\k_2,\k_3)$.  We direct
the $z$ axis along $\k_1$ and introduce the vectors
$\f_2=\k_{2\perp}/\omega_2$ and  $\f_3=\k_{3\perp}/\omega_3$
($|\f_{2,3}|\ll 1$). The $z$ component of the polarization
vectors $\e_i$ can be eliminated owing to the relation $\e_i\k_i=0$
which leads to $e_z=-\e_\perp\k_\perp/\omega$. After that within the
small-angle approximation one can neglect the difference between
the vectors $(\e_{2,3})_\perp$ and the polarization vectors of
photons, propagating along the $z$ axis and having the same
helicities.  Therefore, the amplitudes
$M_{\lambda_1\lambda_2\lambda_3}(\k_1,\k_2,\k_3)$ contain only the
transverse polarization vectors $\e$ and $\e^*$, corresponding to
the positive and negative helicities, respectively. It is sufficient
to calculate three amplitudes, for instance,
$M_{+--}(\k_1,\k_2,\k_3)$ , $M_{+++}(\k_1,\k_2,\k_3)$ and
$M_{++-}(\k_1,\k_2,\k_3)$.
Other amplitudes can be obtained by the substitution
$$
M_{+-+}(\k_1,\k_2,\k_3)=M_{++-}(\k_1,\k_3,\k_2)\, ,
\quad M_{-\lambda_2\lambda_3}(\k_1,\k_2,\k_3)=
M_{+\Lambda_2\Lambda_3}(\k_1,\k_2,\k_3)\,
(\e\leftrightarrow \e^*)\, ,
$$
where $\Lambda$ denotes the helicity opposite to $\lambda$.

It was shown in \cite{LMS1} that the main contribution to the term $M^{(3)}$
is given by the region where $z_1 < 0$ and at least
one of $z_2$ and $z_3$ is positive. Similarly, the main contribution
to the term $M^{(2)}$ is given by the region $z_1 < 0$ and $z_2 > 0$. General
formulae for these terms at arbitrary relations between the
electron mass and the transverse momenta of the final photons
 are presented by Eqs. (9-14) and Eqs. (16-17) in \cite{LMS1}.
Performing an integration by parts in the expression for the term $M^{(3)}$
as it has been done when deriving Eqs. (18-19) in \cite{LMS1}, but
keeping now the terms containing the electron mass, we obtain for the
 amplitude $M$ of the photon splitting
\beq\label{M}
M=M_1+M_2+\delta M \quad ,
\eeq
where $M_1$ corresponds to the contribution to $M^{(3)}$ from the
region
$z_1<0<z_2,\,z_3$, while $M_2$ corresponds to that from two regions
$z_1<z_3<0<z_2$ and $z_1<z_2<0<z_3$. The quantity
$\delta M$ is a sum of the term $M^{(2)}$ and the integrated
terms arising when $M^{(3)}$ is integrated by parts   .

For the term $M_1$, we have
\beqn\label{M1}
\dst
(M_1)_{\lambda_1\lambda_2\lambda_3}
=\fr{{ie^3}}{16\pi^3\omega_1\omega_2\omega_3}
\int\limits_0^{\omega_2}\eps\Q_2\Q_3\, d\eps\int\limits_0^{\infty}dR_1\!
\int\limits_0^{\infty}dR_2\!\int\limits_0^{L}\frac{dR_3}{R_1
R}\times\\
\dst
 \int\!\int d\q_2\,d\q_3 \,
T_{\lambda_1\lambda_2\lambda_3}\, \mbox{e}^{i\Phi}\,
\mbox{Im}\left(\fr{q_2}{q_3}\right)^{2iZ\alpha}\, + \,
(\omega_2\leftrightarrow\omega_3\, ,\,\k_2\leftrightarrow\k_3
\, ,\, \lambda_2\leftrightarrow \lambda_3),\nonumber
\eeqn
where $\q_{2,3}$ are two-dimensional vectors lying in the $xy$ plane,
$\Q_2=\omega_2-\eps\ ,\, \Q_3=\omega_3+\eps$,
$L=R_2\omega_3\Q_2/\omega_2\Q_3$ ,

\beqn\label{Phi}
\dst
\Phi=\biggl[\left(\fr{1}{R}+\fr{1}{R_1}\right)\fr{\vq^2}{2}
+\fr{\eps^2R_2R_3\,\f_{23}^2}{2R} -
\fr{(\Q_2\q_2-\Q_3\q_3,\vd )}{\omega_1}- \\
\dst
\fr{(\omega_3\Q_2R_2-\omega_2\Q_3R_3)}{\omega_1R}(\vq\f_{23})
 -\fr{m^2}{2}(R_1+R)\,\biggr]\, ,\nonumber\\
\dst
R=R_2 -R_3\, ,\quad \f_{23}=\f_2-\f_3\, ,\quad \vq=\q_2+\q_3 \, ,
\quad \vd =\omega_2\f_2+\omega_3\f_3\, . \nonumber
\eeqn
 The function $T$  for different polarizations is:
\beqn\label{T}
\dst
T_{+--}=\fr{8}{R_1R^2}(\e\vq)(\e\vq_2)(\e\vq_3)\, ,\\
\dst
{}\nonumber\\
\dst
T_{+++}=-\fr{4}{R_1R^2}\left(\fr{\Q_2}{\Q_3}+\fr{\Q_3}{\Q_2}\right)
(\e\vq)(\e^*\vq_2)(\e^*\vq_3) +\fr{2m^2\omega_1}{\eps
R}\e^*\left(\fr{\omega_2}{\Q_2}\vq_2-\fr{\omega_3}{\Q_3}\vq_3\right)
\  ,\nonumber\\
\dst
{}\nonumber\\
\dst
T_{++-}=-\fr{4}{R_1R^2}\left(\fr{\Q_2}{\eps}+\fr{\eps}{\Q_2}\right)
(\e\vq)[\, (\e\vq_2)(\e^*\vq_3)\, - iR\,]
+\fr{2m^2\omega_2}{\Q_3}\e\left(\fr{\omega_1}{\Q_2 R}\vq_2-
\fr{\omega_3}{\eps R_1}\vq\right)\ , \nonumber\\
\dst
{}\nonumber\\
\dst
T_{+-+}=\fr{4}{R_1R^2}\left(\fr{\Q_3}{\eps}+\fr{\eps}{\Q_3}\right)
(\e\vq)[\, (\e^*\vq_2)(\e\vq_3)\, - iR\,]
+\fr{2m^2\omega_3}{\Q_2}\e\left(\fr{\omega_1}{\Q_3 R}\vq_3+
\fr{\omega_2}{\eps R_1}\vq\right)\ , \nonumber
\eeqn
where $\vq_2=\vq+\eps R_2\f_{23}$ and  $\vq_3=\vq+\eps R_3\f_{23}$.

The term $M_2$ is given by

\beqn\label{M2}
 \dst
(M_2)_{\lambda_1\lambda_2\lambda_3}=
 \fr{{ie^3}}{16\pi^3\omega_1\omega_2\omega_3}
\int\limits_0^{\omega_2}\eps\Q_2\Q_3\, d\eps\int\limits_0^{\infty}dR_1\!
\int\limits_0^{\infty}dR_2\!\int\limits_0^{\tilde
L}\frac{dR_3}{R_2\tilde{R}}\times\\
 \dst
  \int\!\int d\q_2\,d\q_3\,
\tilde T_{\lambda_1\lambda_2\lambda_3}\,\mbox{e}^{i\tilde\Phi}\,\mbox{Im}
 \left(\fr{q_2}{q_3}\right)^{2iZ\alpha}\, + \,
(\omega_2\leftrightarrow\omega_3\, ,\,\k_2\leftrightarrow\k_3
\, ,\, \lambda_2\leftrightarrow \lambda_3) \ ,\nonumber
 \eeqn
 where $\tilde L=R_1\omega_3\Q_2/\omega_1\eps$ , $\tilde{R}=R_1+R_3$ ,
and the functions $\tilde\Phi$ and $\tilde T$ can be obtained from
$\Phi$ and $T$ in \eq{Phi},\eq{T}
by the substitutions
\beqn\label{subs}
\dst
\q_{2,3}\rightarrow -\q_{2,3}\quad  , \quad
\omega_1\leftrightarrow \omega_2\quad , \quad \omega_3\rightarrow
-\omega_3\quad ,\quad \Q_3\leftrightarrow\eps \quad , \nonumber\\
\dst
R_1\leftrightarrow R_2\quad ,\quad R_3\rightarrow -R_3\quad ,
\quad \f_{23} \leftrightarrow -\f_3\quad , \quad \f_2\rightarrow
-\f_2\, ,
\eeqn
so that
$$
T_{+--}\rightarrow \tilde T_{+--}\, ,\quad
T_{+-+}\rightarrow \tilde T_{+-+}\, ,\quad
T_{+++}\rightarrow \tilde T_{++-}(\e\leftrightarrow\e^*)\, ,\quad
T_{++-}\rightarrow \tilde T_{+++}(\e\leftrightarrow\e^*)\, .
$$

For the last contribution to the amplitude $M$ \eq{M} we obtain
\beqn\label{delM}
\dst
(\delta M)_{\lambda_1\lambda_2\lambda_3}
=-\fr{{e^3}}{4\pi^3} \int\limits_0^{\infty}\! dR_1
\int\limits_0^{\infty}\!\fr{dR_2}{R_1^2 R_2} \int\!\!\int
d\q_2\,d\q_3 \, F_{\lambda_1\lambda_2\lambda_3}
\,\mbox{Im}\left(\fr{q_2}{q_3}\right)^{2iZ\alpha} \quad ,
\eeqn
where
\beqn\label{F}
\dst
F_{+--}=0\quad ,\quad
F_{+-+}=(\e\vq)\left[
\int\limits_{-\omega_3}^{\omega_2}\!
d\eps\, \fr{\Q_2\Q_3^2}{\omega_1^2\eps}\mbox{e}^{i\psi_1}-
\int\limits_{0}^{\omega_2}\!
d\eps\, \fr{\Q_2\eps^2}{\omega_2^2\Q_3}\mbox{e}^{i\psi_2}\right]\, ,
\nonumber\\\
\dst\
F_{+++}=(\e^*\vq)\int\limits_{0}^{\omega_2}\!\
d\eps\, \fr{\eps\Q_2^2}{\omega_2^2\Q_3}\mbox{e}^{i\psi_2}+\
(\omega_2\leftrightarrow\omega_3\, ,\,\f_2\leftrightarrow\f_3)\quad ,\
\nonumber\\\
\dst\
F_{++-}=F_{+-+}(\omega_2\leftrightarrow\omega_3\,\
,\,\f_2\leftrightarrow\f_3)\quad , \\\
\dst\
{}\nonumber\\\
\dst\
\psi_1=(\fr{1}{R_1}+\fr{1}{R_2})\fr{\vq^2}{2}\
+\fr{\omega_2\omega_3\Q_2\Q_3}{2\omega_1^2}\f_{23}^2R_2 -\
\fr{(\Q_2\q_2-\Q_3\q_3,\vd )}{\omega_1}-\fr{m^2}{2}(R_1+R_2)\quad\
,\nonumber\
\dst\
{}\nonumber\\\
\dst\
\psi_2=(\fr{1}{R_1}+\fr{1}{R_2})\fr{\vq^2}{2}\
-\fr{\omega_1\omega_3\Q_2\eps}{2\omega_2^2}\f_{3}^2R_2 -\
\fr{(\Q_2\q_2-\eps\q_3,\vd )}{\omega_2}-\fr{m^2}{2}(R_1+R_2)\quad\
.\nonumber\
\eeqn

The expressions derived are still rather complicated and require
further transformations. Let us pass from the variables $\q_2,\,\q_3$
to $\vq=\q_2+\q_3$ and $\q=\q_2-\q_3$ and use the
identity (see  Eqs. (22,23) in \cite{LMS1})

\beq
\int\fr{d\q}{\vq^2}\,\exp(-\fr{i}{2}\q\vd)\,
\mbox{Im}\left(\fr{|\q+\vq|}{|\q-\vq|}\right)^{2iZ\alpha}
=
\int\fr{d\q}{\vd^2}\,\exp(-\fr{i}{2}\q\vq)\,
\mbox{Im}\left(\fr{|\q+\vd|}{|\q-\vd|}\right)^{2iZ\alpha}\ .
\eeq

Additionally, the parametrization

\beq
\exp(i\fr{\vq^2}{2R_1})=iR_1\int \fr{d\vx}{2\pi}\exp(-i\fr{R_1
\vx^2}{2}-i\vq\vx)\ ,
\eeq
where $\vx$ is a two-dimensional vector, is used to calculate the
term $M_1$.

After that it is easy to take first the integrals over
$R_1$, $\vq$, and then over $R_2$ and $R_3$.
The calculation of the terms $M_2$ and $\delta M$
is carried out analogously. Performing, finally,
 the shift $\vx\rightarrow \vx-\q/2$, we obtain only two kinds of integrals
 over $\q$ to be taken:

\begin{eqnarray}
\label{G}
\dst
{\mathbf G} =m^2\int \frac{d{\mathbf q}}{2\pi }\,\left( \frac{{\mathbf q}_{+}
}{q_{+}^2}-\frac{{\mathbf q}_{-}}{q_{-}^2}\right)
\frac 1{[m^2+({\mathbf x-q/}2)^2]^2}\mbox{Re}\left(
\frac{q_{+}}{q_{\_}}\right) ^{2iZ\alpha }=\\ \dst \cos \,
\tau\,\left( \frac{{\mathbf x+\Delta /}2}{c_{+}}-\frac{{\mathbf
x-\Delta /}2}{c_{-}}\right) {\cal F}_1+\sin \, \tau\,\left(
\frac{{\mathbf x+\Delta /}2}{c_{+}}+ \frac{{\mathbf x-\Delta
/}2}{c_{-}}\right) {\cal F}_2\ ,\nonumber\\
\dst
{}\nonumber \\
\dst
G_1 =2 m^2\int \frac{d{\mathbf q}}{2\pi }\,\left(
\frac{{\mathbf eq}_{+}}{
q_{+}^2}-\frac{{\mathbf eq}_{-}}{q_{-}^2}\right)
 \frac{{\mathbf e}^{*}({\mathbf
x-q/}2)}{[m^2+({\mathbf x-q/}2)^2]^2}\mbox{Re}\left(
\frac{q_{+}}{q_{\_}}\right) ^{2iZ\alpha }= \nonumber \\
\dst
m^2\cos \, \tau\,\left( \frac 1{c_{-}}-\frac 1{c_{+}}\right) {\cal F}_1+
\sin \, \tau\,\left( 1-\frac{m^2}{c_{+}}-\frac{m^2}{c_{-}}\right)
{\cal F}_2\nonumber\ ,
\end{eqnarray}
where the  notation
\[
\q_{\pm}=\q \pm \vd \ , \
c_{\pm }=m^2+\left(
{\mathbf x\pm \Delta /}2\right) ^2\;,\;a=\sqrt{\fr{m^2\vd^2 }{
c_{+}c_{-}}}\;,\;\tau=Z\alpha \ln \left(
\frac{c_{+}}{c_{-}}\right) \
\]
  is introduced.  The functions ${\cal F}_1$ and ${\cal F}_2$ are given by
\begin{eqnarray}
\label{f12}
\dst
{\cal F}_1 =\int_0^\infty
\frac{dx}{\left( 1+x^2\right) ^{3/2}}\cos \left( 2Z\alpha
\,\mbox{arcsinh}(ax)\right) \ , \\
\dst
{\cal F}_2 =\int_0^\infty \frac{dx}{\left(
1+x^2\right) ^{3/2}}\sin \left( 2Z\alpha \,\mbox{arcsinh}(ax)\right)
\frac{ax}{\sqrt{1+a^2x^2}}\ .\nonumber
\end{eqnarray}
We explain in Appendix I how the integrals in \eq{G} have been actually taken.
The representation \eq{G} is very useful since a dependence of
  the functions ${\cal F}_1$ and ${\cal F}_2$ on $\vx$ is expressed by the
single variable $a$. This greatly simplifies numerical calculations.

Finally, we obtain for the helicity amplitudes of photon splitting
\begin{eqnarray}
\label{final}
\dst
M_{+--} =N\int
d{\mathbf x\,}\left( {\mathbf eG}\right) \int_0^{\omega _2}
\frac{d\varepsilon }A\kappa _2\left[ -\left( {\mathbf ea}\right) \left( \frac
\varepsilon {\left( {\mathbf e}^{*}{\mathbf f}_3\right) }+\frac{\kappa _3}{
\left( {\mathbf e}^{*}{\mathbf f}_{23}\right) }\right) +\right.  \\
\dst
\left. m^2\omega _3\kappa _2\left( \frac{\kappa _3\left( {\mathbf ef}_{23}
\right) }{\omega_1{\cal D}_1\left( {\mathbf e}^{*}{\mathbf f}
_{23}\right) }-\frac{\varepsilon \left( {\mathbf ef}_3\right) }{\omega_2
{\cal D}_3\left( {\mathbf e}^{*}{\mathbf f}_3\right) }\right) \right]
+\left(
\begin{array}{c}
\omega_2\leftrightarrow \omega_3 \\
{\mathbf f}_2\leftrightarrow {\mathbf f}_3
\end{array}
\right) \ ,  \nonumber\\
\dst
{}\nonumber\\
\dst
M_{+++} =N\int d{\mathbf x\,}\int_0^{\omega _2}\frac{d\varepsilon
}{2A}\left[ \frac{\left( {\mathbf eG}\right) \left( \kappa _2^2+\kappa
_3^2\right) }{\left( {\mathbf ef}_{23}\right) }\left( \left(
{\mathbf e}^*{\mathbf a}\right) -\frac{ m^2\omega _3\kappa _2\left(
{\mathbf e}^{*}{\mathbf f}_{23}\right) }{\omega _1
{\cal D}_1}\right) \right. + \nonumber\\
\dst
\frac{\omega _1\omega _3\kappa _2G_1{\mathbf e}^{*}\left( {\mathbf a+f}
_{23}\omega _2\kappa _3/\omega _1\right) }{{\cal D}_1}+\frac{\omega
_3\kappa _2\left( {\mathbf e}^{*}{\mathbf G}\right) }{\omega _2{\cal D}_3}
[\varepsilon \omega _3A+ \nonumber\\
\dst
2\varepsilon \left( \kappa _2^2+\kappa _3^2\right) \left( {\mathbf e}^{*}
{\mathbf f}_3\right) \left( {\mathbf ea}\right) +m^2(2\varepsilon \kappa
_2-\omega _1\omega _2)] \nonumber\\
\dst
\left. -\frac{\omega _1\omega _3\kappa _2G_1\left( {\mathbf e}^{*}{\mathbf c}
\right) }{{\cal D}_3}\right] +\left(
\begin{array}{c}
\omega _2\leftrightarrow \omega _3 \\
{\mathbf f}_2\leftrightarrow {\mathbf f}_3
\end{array}  \right)\ , \nonumber\\
\dst
{}\nonumber\\
\dst
M_{++-} =N\int d{\mathbf x\,}\left\{ \int_0^{\omega _2}\frac{d\varepsilon }
{2A}\left[ \frac{\kappa _2\omega _3\left( {\mathbf eG}\right)
}{\omega _1 {\cal D}_1}[\kappa _3\left( \kappa _2-\varepsilon
\right) A-\right.  \right.  \nonumber\\
\dst
 2\kappa _3\left( \kappa_2^2+\varepsilon ^2\right) \left( {\mathbf ef}
 _{23}\right) \left(
{\mathbf e}^{*}{\mathbf a}\right) +m^2\left( \omega _1\omega _2-2\kappa
_2\kappa _3\right) ]+\frac{\kappa _2\omega _2\omega _3G_1\left(
{\mathbf eb}\right) }{{\cal D}_1}+ \nonumber\\
\dst
\left. \frac{\left( \kappa
_2^2+\varepsilon ^2\right) \left( {\mathbf e}^{*} {\mathbf G}\right)
}{\left( {\mathbf e}^{*}{\mathbf f}_3\right) }\left( \left(
{\mathbf ea}\right) +\frac{m^2\omega _3\kappa _2\left( {\mathbf ef}_3\right) }{
\omega _2{\cal D}_3}\right) -\frac{\omega _2\omega _3\kappa _2G_1{\mathbf e
}\left( {\mathbf a+f}_3\omega _1\varepsilon /\omega _2\right) }{{\cal D}_3}
\right] + \nonumber\\
\dst
{}\nonumber\\
\dst
\int_{-\omega _3}^0d\varepsilon \frac{\omega _2\kappa _3}{2B}\left[
\frac{ \left( {\mathbf eG}\right) }{\omega _1{\cal D}_1}[-\left(
\kappa _2^2+\varepsilon \kappa _3\right) B+2\kappa _3\left( \kappa
_2^2+\varepsilon ^2\right) \left(
{\mathbf e}^{*}{\mathbf f}_{23}\right) \left( {\mathbf eb} \right)
+\right.  \nonumber\\
\dst
m^2\left( \omega _1\omega _2-2\kappa _2\kappa
_3\right) ]+\frac{\omega _2G_1\left( {\mathbf eb}\right)
}{{\cal D}_1}+\frac{\left( {\mathbf eG} \right) }{\omega
_3{\cal D}_2}[-\omega _2\kappa _3B+ \nonumber\\
\dst
\left. \left.
2\kappa _3\left( \kappa _2^2+\varepsilon ^2\right) \left(
{\mathbf e}^{*}{\mathbf f}_2\right) \left( {\mathbf eb}\right) +m^2\left(
\omega _2\omega _3-2\varepsilon \kappa _3\right) ]+\frac{\omega _2G_1\left(
{\mathbf eb}\right) }{{\cal D}_2}\right] \right\}\nonumber\ ,
\end{eqnarray}

where the following notation is used

\beqn
\dst
\A=\vx-\vd/2+\Q_2\f_2\quad , \quad \B=\vx+\vd/2-\Q_3\f_3\quad , \quad
\C=\vx+\vd/2-\eps\f_{23}\quad ,\\
\dst
N=\frac{8e^3Z\alpha }{\pi ^2\Delta^2\omega _1\omega _2\omega_3}
\quad ,\quad
A=m^2+\A^2\quad,\quad B=m^2+\B^2\nonumber\ ,\\
\dst
{\cal D}_1=\left(\vx+\fr{\Q_2-\Q_3}{2\omega_1}\,\vd
\right)^2-\fr{\omega_2\omega_3\Q_2\Q_3}{\omega_1^2}\f_{23}^2-i0\quad
, \\
\dst
{\cal D}_2=\left(\vx-\fr{\Q_3+\epsilon}{2\omega_3}\,\vd
\right)^2-\fr{\omega_1\omega_2\Q_3\epsilon}{\omega_3^2}\f_{2}^2\quad
,\nonumber\\
\dst
{\cal D}_3=\left(\vx+\fr{\Q_2-\epsilon}{2\omega_2}\,\vd
\right)^2+\fr{\omega_1\omega_3\Q_2\epsilon}{\omega_2^2}\f_{3}^2\quad
.\nonumber
\end{eqnarray}

As was pointed out in \cite{LMS1}, the quantity $\Delta^2$ in
the coefficient $N$ should be interpreted as the squared total
momentum transfer:
\beq\label{Delta}
\Delta^2=(\k_2+\k_3-\k_1)^2=\vd^2+\Delta_z^2=\,
(\k_{2\perp}+\k_{3\perp}\,)^2+\frac{1}{4}
\left(\frac{\k_{2\perp}\,^2}{\omega_2}+
\frac{\k_{3\perp}\,^2}{\omega_3}\right)^2\, .
\eeq

Since the functions $\mathbf G$ and $G_1$ are independent of
the energy $\eps$, the integrands in \eq{final} are rational functions
of $\eps$, where all the denominators are quadratic forms of this
variable. Therefore, the integrals over $\eps$ can be expressed
via elementary functions. Resulting formulae being rather cumbersome
are not presented here explicitly ( some details of this integration
are described in Appendix II). Performing the integration over $\eps$
in \eq{final},we obtain in fact a twofold integral over $\vx$ for the
amplitude of photon splitting since for a given $\al$, the functions
 ${\cal F}_1$ and ${\cal F}_2$ (see
\eq{f12}) can be tabulated separately.
In the limit $m\rightarrow 0$ the amplitudes  \eq{final} coincides with those
obtained previously in \cite{LMS2}. Additionally, it has been checked
numerically that in the limit $\al\rightarrow 0$ our results
 \eq{final} agree
with those, obtained in \cite{JMO,S} in the Born approximation.

\section{Cross section}

In the small-angle approximation ($|\f_2|,|\f_3|\ll 1$) the cross
section of photon splitting has the form
\beq \label{cross1}
d\sigma=|M|^2\,\fr{dx \, d\k_{2\perp}\, d\k_{3\perp}\,
}{2^8 \pi^5\omega_1^2x(1-x)} \quad,
\eeq
where $x=\omega_2/\omega_1$, so that $\omega_3=\omega_1(1-x)$.
As was mentioned above, in the general case of a screened Coulomb
potential the lowest in $\al$ (Born) part of the amplitude should be
multiplied by the atomic form factor. The  Moli\`ere representation
 \cite{M} for this form factor reads
\beq\label{FF}
1-F(\Delta^2)=\Delta^2\sum_{i=1}^{3}\,\fr{\alpha_{i}}{\Delta^2+\beta_{i}^2}
\, ,
\eeq
where
\beqn
\label{coef}
\dst
\alpha_{1}=0.1\quad , \quad \alpha_{2}=0.55 \quad,\quad
\alpha_{3}=0.35 \quad, \quad \beta_{i}=\beta_0 b_i \quad,
\\
\dst \nonumber
\\
\dst \nonumber
b_{1}=6 \quad, \quad b_{2}=1.2 \quad , \quad b_{3}=0.3 \quad ,
\quad \beta_0=\, mZ^{1/3}/121\quad .
\eeqn
 To illustrate a magnitude of the Coulomb corrections,
 the exact and Born differential cross sections
 $d\sigma/dxd\k_{2\perp}d\k_{3\perp}$   are plotted in Fig.1 depending
on $|k_{2\perp}|/m$ for  $|\k_{3\perp}|=2m$ and the azimuth
angle (the angle between the vectors $\k_{2\perp}$ and
$\k_{3\perp}$) $\phi=0,\,\pi$.
The calculations were performed for a screened
Coulomb potential at $x=0.1$, $\omega_1=1$GeV. The value  $Z=83$
(bismuth) was chosen since bismuth atoms determine
the cross section of photon splitting in the experiment
\cite{Budker}. A wide peak for azimuth angle $\phi=\pi$ is due to
small momentum transfer $\Delta$. There is a narrow notch
in the middle of this peak ( at $|k_{2\perp}|/m=2$ ) where the
condition $\vd_\perp=\k_{2\perp}+\k_{3\perp}=0$ is fulfilled. The
width of the notch is about $\mbox{max}(\Delta_z/m,\beta_0/m)$.
Recall that $\Delta_z$ is the longitudinal component of the
momentum transfer defined by \eq{Delta}, and $\beta_0$ \eq{coef}
characterizes the effect of screening. In our example  $\beta_0$  is larger
than $\Delta_z$ , so the width of the notch is roughly
$\beta_0/m=3.6\cdot 10^{-2}$.
Let us note that for $\omega_1\gg m$ the differential cross
section, expressed in terms of $\k_{2\perp}$,
$\k_{3\perp}$, $x$ and $\omega_1$ depends on the energy $\omega_1$
only via $\Delta_z$. Due to this, the
differential cross section is independent of $\omega_1$ outside
the notch vicinity. The behavior of the cross section at small $\Delta$
is determined by the Born amplitude which is proportional to
$1/\Delta$ in this case. That is why the  exact
and Born cross sections coincide within peak region.
Outside this region the
Coulomb corrections essentially modify the cross section.
The points of discontinuous slope on the curves in Fig.1 are related to the
threshold conditions
for real electron-positron pair production by two photons with
the momenta $\k_2$ and $\k_3$:

\beq\label{thr} (k_2+k_3)^2 =
\omega_2\omega_3\f_{23}^2=4m^2\ .
\eeq

In Figs. 2-4 the differential cross section $m^2 \sigma_0^{-1}
d\sigma/dx\, d\k_{2\perp}$ is shown depending on $k_{2\perp}/m$
for a screened Coulomb potential at $\omega_1/m=1000$,
$Z=83$ and different $x$,
$$
\sigma_0=\fr{\alpha^3(\al)^2}{4\pi^2m^2}=0.782\cdot 10^{-9} \, Z^2
\mbox{ b}\ .
$$
Solid curves represent the exact cross sections, and the dashed
curves give the Born results. The cross section exhibits a thresholdlike
behavior in the vicinity of the point $k_{2\perp}=k_{th}=2\sqrt{x(1-x)}m$,
where both conditions
$\vd=\k_{2\perp}+\k_{3\perp}=0$ and \eq{thr} hold. Under these conditions
the peak in the cross section $d\sigma/dxd\k_{2\perp}d\k_{3\perp}$
seats on the boundary of the kinematic region where
  real electron-positron pair production by two photons with the
momenta $\k_2$ and $\k_3$ is possible.
The cross sections
$d\sigma/dx\, d\k_{2\perp}$ drop rapidly for $k_{2\perp}\gg m$
($\propto 1/k_2^4$).
The dotted curves in Figs. 2-4 show the difference between the Born
and exact cross sections, i.e. they give the Coulomb corrections
taken with the opposite sign. Again, as is seen from  Figs.2-4, the
Coulomb corrections to the cross section integrated over
$\k_{3\perp}$ noticeably diminish the magnitude of the cross section.
Above the threshold ($k_{2\perp}>k_{th}$) this difference reaches
tens per cent while below the threshold the exact cross section is
several times smaller than the Born one. It can be explained as
follows.  Above the threshold the main contribution to the cross
section $d\sigma/dx\, d\k_{2\perp}$ is given by the integration
region where $\mbox{max}(\k_{2\perp}^2/\omega_1,\beta_0)\ll\Delta\ll
k_{2\perp}$.  As a result, the Born cross section is logarithmically
amplified as compared to the Coulomb corrections. Far below the
threshold where $k_{2\perp}\ll k_{th}$, it follows from the condition
$\Delta\ll k_{2\perp}$ that $k_{3\perp}\approx k_{2\perp} \ll m$, and
the amplitude is suppressed as a power of $k_{2\perp}^2/m^2$.
Therefore, below the threshold the main contribution to the cross
section $d\sigma/dx\, d\k_{2\perp}$ is given by the region
$k_{3\perp}\sim m$, where the exact in $\al$ amplitude drastically
differs from the Born one.
 It is seen from Figs.3,4 that in accordance with previous discussion a
 position of the peak is the same for $x=0.2$ and $x=0.8$. Nevertheless, the
magnitudes of these two cross sections are significantly different
, especially below the threshold. The explanation is the
following. For the region $k_{3\perp}\sim m$, which makes the main
contribution to the cross section $d\sigma/dx\, d\k_{2\perp}$ below
the threshold, and $k_{2\perp}\ll m$ the invariant $s=(k_2+k_3)^2\sim
m^2 x/(1-x)$. So, $s\ll m^2$ at $x\ll 1$ which leads to the
suppression of the cross section below the threshold (see Fig. 3).
This is not the case for $1-x\ll 1$ (see Fig. 5). However, the cross
section $d\sigma/dx$ for $x=0.2$ should coincide with that for
$x=0.8$ and this was checked numerically.

 Let us consider now the magnitude of the Coulomb corrections to the
cross section  $d\sigma/dx$ integrated over the transverse momenta of
both final photons. The main contribution to this cross section is
given by the region where $|\k_{2\perp}|,\,|\k_{3\perp}| \sim m$.
 The Born contribution to $d\sigma/dx$ contains
large logarithm resulting from the integration over small
momentum transfer region $\mbox{max}(\beta_0,m^2/\omega_1)\ll\Delta\ll m$.
For $\beta_0\gg m^2/\omega_1$ the cross section $d\sigma/dx$
is independent of $\omega_1$, while for $\beta_0\ll m^2/\omega_1$ it
slowly grows (as $\ln \omega_1/m$) when $\omega_1$ increases. Since the Coulomb
corrections to $d\sigma/dx$ are determined by the region of momentum transfer
$\Delta\sim m$, they do not depend on $\omega_1$ for
$\omega_1\gg m$ . They also are
insensitive to the effect of screening. In Fig. 5 the
exact (solid curve) and the Born (dashed curve) cross sections
$\sigma_0^{-1} d\sigma/dx$ are plotted as functions of $x$ for
$\omega_1/m=1000$, and $Z=83$. As it should be, the curves are
symmetric with respect to the replacement $x\rightarrow 1-x$.
Dotted curve shows the Coulomb corrections
taken with the opposite sign. Note that their dependence
on $x$ is very weak. If $x\rightarrow 0$ or $x\rightarrow 1$
then the cross section $d\sigma/dx$  increases rapidly.  However,
the cross section $d\sigma/dx$ should vanish
at $x=0$ and $x=1$ due to the gauge invariance of QED. Actually,
the cross section $d\sigma/dx$ begins to decrease very close to the
$x$ interval end points ($\delta x\sim m^2/\omega_1^2$). Therefore,
the contribution from these  $x$-range to the total cross section
$\sigma$ is negligible. In our example ( $Z=83$, $\omega_1/m=1000$ ),
the exact result for $\sigma$ is  $3.9\cdot 10^{-4}$b while
the Born approximation gives  $4.8\cdot 10^{-4}$b, the difference
being 23\%.

In Fig. 6 the Coulomb corrections
$d\sigma_c/dx$ divided by $\sigma_0$ are shown as a function
of $Z$ for $x=0.7$.  Since their dependence on $x$ is rather weak (
see Fig.  5), this curve allows one to estimate the magnitude of the
Coulomb corrections for any $x$. It can be seen from Fig. 6 that
starting from $Z\approx 30$ the dependence of $\sigma_0^{-1} d\sigma_c/dx$
on $Z$ is almost linear and this quantity is
not described by its lowest in $\al$ approximation ($\propto
(\al)^2$), so that higher order terms are important.

Thus, the process of photon splitting can be adequately described
only with the Coulomb corrections taken into account.
At large $Z$ their contribution is always essential, though
the magnitude of the Coulomb corrections depends on the type of the
cross section, kinematic conditions, and $Z$.
 Indeed, the predictions based on the exact in $\al$ cross
section are in agreement with the preliminary experimental data of
\cite{Budker}, while the Born results are noticeably different.

\section*{Appendix I}

In this Appendix we describe the calculation of the integrals in
\eq{G}.
Let us consider the integral
\beq
\label{GA}
{\mathbf G} =m^2\int \frac{d{\mathbf q}}{2\pi }\,\left( \frac{{\mathbf q}_{+}
}{q_{+}^2}-\frac{{\mathbf q}_{-}}{q_{-}^2}\right) \frac
1{[m^2+({\mathbf x-q/} 2)^2]^2}\mbox{Re}\left(
\frac{q_{+}}{q_{\_}}\right) ^{2iZ\alpha } \ , \eeq where $\q_{\pm}=\q
\pm \vd$. To transform this integral, we multiply the integrand in
\eq{GA} by
$$
1\equiv\int\limits_{-1}^{1}dy\,\delta
\left(y-\fr{2\q\vd}{\q^2+\vd^2} \right)=
(\q^2+\vd^2)\int\limits_{-1}^{1}\fr{dy}{|y|}\,
\delta((\q-\vd /y)^2 -\vd^2(1/y^2-1))\quad ,
$$
change the order of integration over $\q$ and $y$ and make the
shift $\q\rightarrow \q+\vd/y$. After that the integral over $q$
becomes trivial, and the integral over the angle of $\q$
can be easily taken by means of the residue technique.
As a result, we obtain
\begin{eqnarray} \dst
{\mathbf G} =-m^2{\mathbf \Delta }^2\int_{-1}^1dy\;
\mbox{Re}\left( \frac{1+y}{1-y}\right) ^{iZ\alpha }
\!\!\!\left( \,y\,{\mathbf x-\Delta }/2\right)
\times  \\
\dst
\left[ m^2{\mathbf \Delta }^2+\left( {\mathbf x\Delta }\right) ^2-2{\mathbf
x\Delta }Cy+\left( C^2-m^2{\mathbf \Delta }^2\right) y^2\right]
^{-3/2},\nonumber \\
\dst
C =m^2+{\mathbf x}^2+{\mathbf \Delta }^2/4\ .\nonumber
\end{eqnarray}
Let us perform the substitution $y=\mbox{tanh}s$ and then make the
shift $s\rightarrow s+(1/2) \ln c_+/c_-$, where $c_{\pm }=m^2+\left(
{\mathbf x\pm \Delta /}2\right) ^2$. Finally, we come to the
expression for ${\mathbf G}$ in \eq{G} with ${\cal F}_1$
 and  ${\cal F}_2$ in the form of
\begin{eqnarray*}
\dst
{\cal F}_1 =a^2\int_0^\infty ds\frac{\cosh \,s\,\cos \left( 2Z\alpha s\right)
 }{%
\left( \sinh ^2s+a^2\right) ^{3/2}} \\
\dst
{\cal F}_2 =a^2\int_0^\infty ds\frac{\sinh \,s\,\sin \left( 2Z\alpha s\right)
}{%
\left( \sinh ^2s+a^2\right) ^{3/2}}\ ,
\end{eqnarray*}
where $a^2=m^2\vd^2/c_+c_-$.
Making in these formulae the substitution $\sinh s=ax$, we obtain
the form \eq{f12}. The quantity $G_1$ can be transformed in the same
way.

\section*{Appendix II}

A general form of the integral over the energy $\eps$ reads
\beq
\label{basic}
\int_0^\omega d\varepsilon \frac{{\cal Q}}{{\cal AB}}\quad ,
\eeq
where
\[
{\cal A}=A_2\varepsilon ^2+A_1\varepsilon +1,\ {\cal B=}B_2\varepsilon
^2+B_1\varepsilon +1\quad ,
\]
and ${\cal Q}$ is the fifth-degree polinomial in $\eps$.
To perform the integration, it is convenient to use the
representation
\begin{eqnarray*}
\frac{\varepsilon ^n}{{\cal AB}} &=&\frac{\delta _{n4}}{A_2B_2}+\delta
_{n5}\left( \frac \varepsilon {A_2B_2}-\frac{A_1B_2+A_2B_1}{\left(
A_2B_2\right) ^2}\right) + \\
&&\frac{p_1^{(n)}\varepsilon +p_0^{(n)}}{{\cal A}}-\frac{%
q_1^{(n)}\varepsilon +q_0^{(n)}}{{\cal B}}\quad ,
\end{eqnarray*}
where the coefficients $p_i^{(n)}$ and $q_i^{(n)}$ satisfy the
recurrence relations
\begin{eqnarray*} p_1^{(n)}
&=&p_0^{(n-1)}-\frac{A_1}{A_2}p_1^{(n-1)},\ p_0^{(n)}=-\frac{%
p_1^{(n-1)}}{A_2} \\
q_1^{(n)}
&=&q_0^{(n-1)}-\frac{B_1}{B_2}q_1^{(n-1)},\ q_0^{(n)}=-\frac{%
q_1^{(n-1)}}{B_2}\ ,
\end{eqnarray*}
\begin{eqnarray*}
p_1^{(0)} &=&\frac{A_2V}W,\ p_0^{(0)}=\frac{A_1V-A_2(A_2-B_2)}W, \\
q_1^{(0)} &=&\frac{B_2V}W,\ q_0^{(0)}=\frac{B_1V-B_2(A_2-B_2)}W, \\
V &=&A_2B_1-A_1B_2,\ W=(A_1-B_1)V-(A_2-B_2)^2
\end{eqnarray*}

Thus, the calculation of the integral \eq{basic} reduces to the
calculation of standard integral
\beq
\label{basic1}
\int_0^\omega d\varepsilon \frac{p_1\eps+p_0}
{A_2\varepsilon ^2+A_1\varepsilon +1}\quad ,
\eeq
expressed via the elementary functions.

This is the algorithm used in our numerical calculations.

\newpage \begin{center} Figure captions \end{center}

Fig. 1. Differential cross section
$d\sigma/dxd\k_{2\perp}d\k_{3\perp}$ vs $|\k_{2\perp}|/m$
 in a screened Coulomb potential for different
azimuth angle $\phi$ between vectors $k_{2\perp}$ and $k_{3\perp}$;
$Z=83$, $x=0.1$, $\omega_1=1$GeV, $k_{3\perp}=2m$.  The dashed
curve (Born approximation) and the solid curve (exact cross section)
correspond to $\phi=\pi$. The dash-dotted curve (Born approximation)
and the dotted curve (exact cross section) correspond to
$\phi=0$.

Fig. 2.  $m^2\sigma_0^{-1} d\sigma/dxd\k_{2\perp}$ vs
$|\k_{2\perp}|/m$ for a screened Coulomb potential,
$\omega_1/m=1000$, $x=0.5$, $Z=83$, $\sigma_0$ is given in the text.
The dashed curve corresponds to the Born approximation, the solid
curve gives the exact result, and the dotted curve shows the
difference between the Born cross section and the exact one.

Fig. 3. Same as Fig. 2 but for $x=0.2$ .

Fig. 4. Same as Fig. 2 but for $x=0.8$ .

Fig. 5. The dependence of $\sigma_0^{-1}d\sigma/dx$ on $x$ for a
screened Coulomb potential, $\omega_1/m=1000$, $Z=83$.  The
dashed curve corresponds to the Born approximation, the solid one
gives the exact result, and the dotted curve shows the difference
between the Born cross section and the exact one.

Fig. 6. The dependence of the Coulomb corrections $\sigma_0^{-1}
d\sigma_C/dx$ on $Z$ for $x=0.7$.

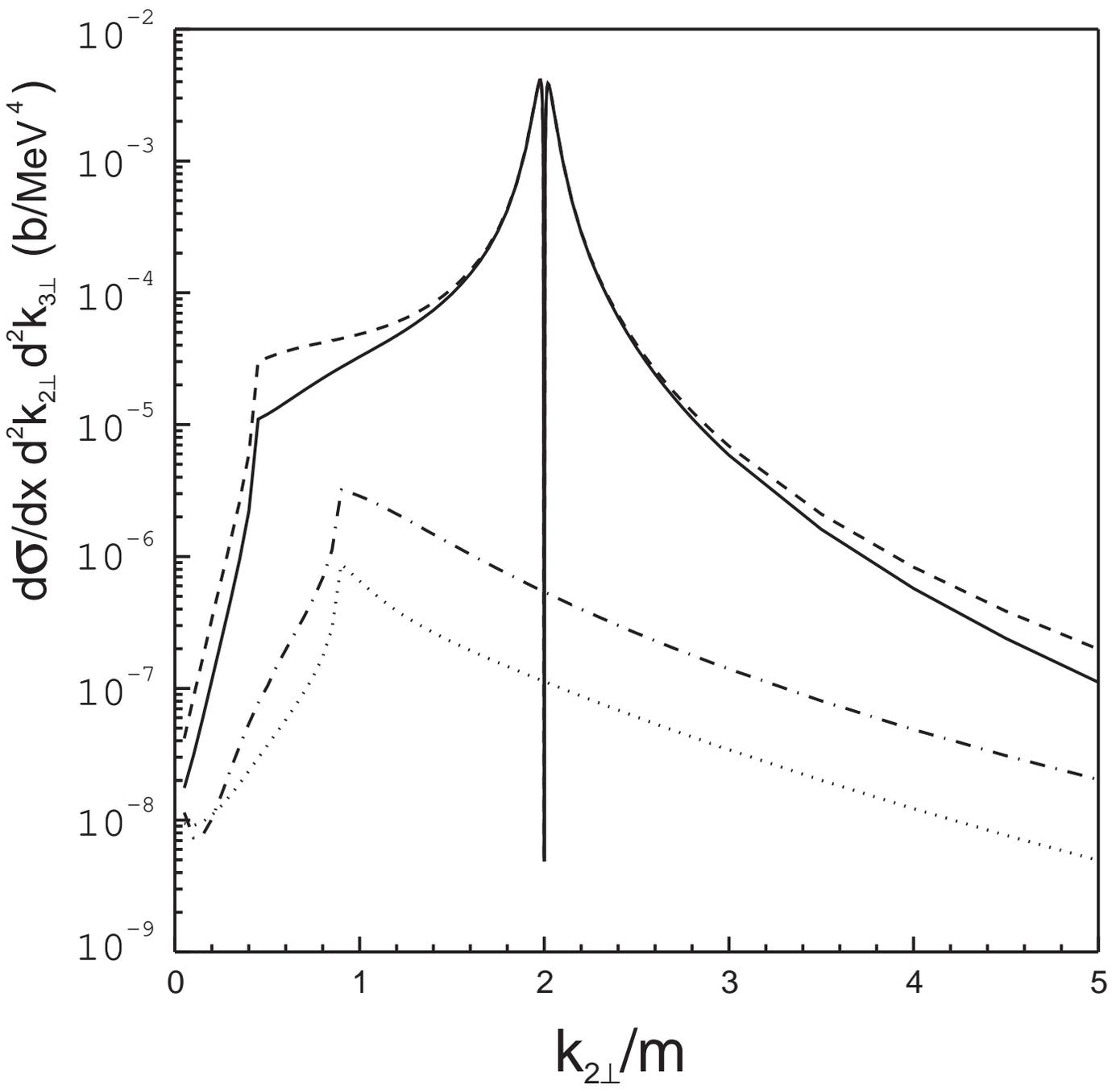
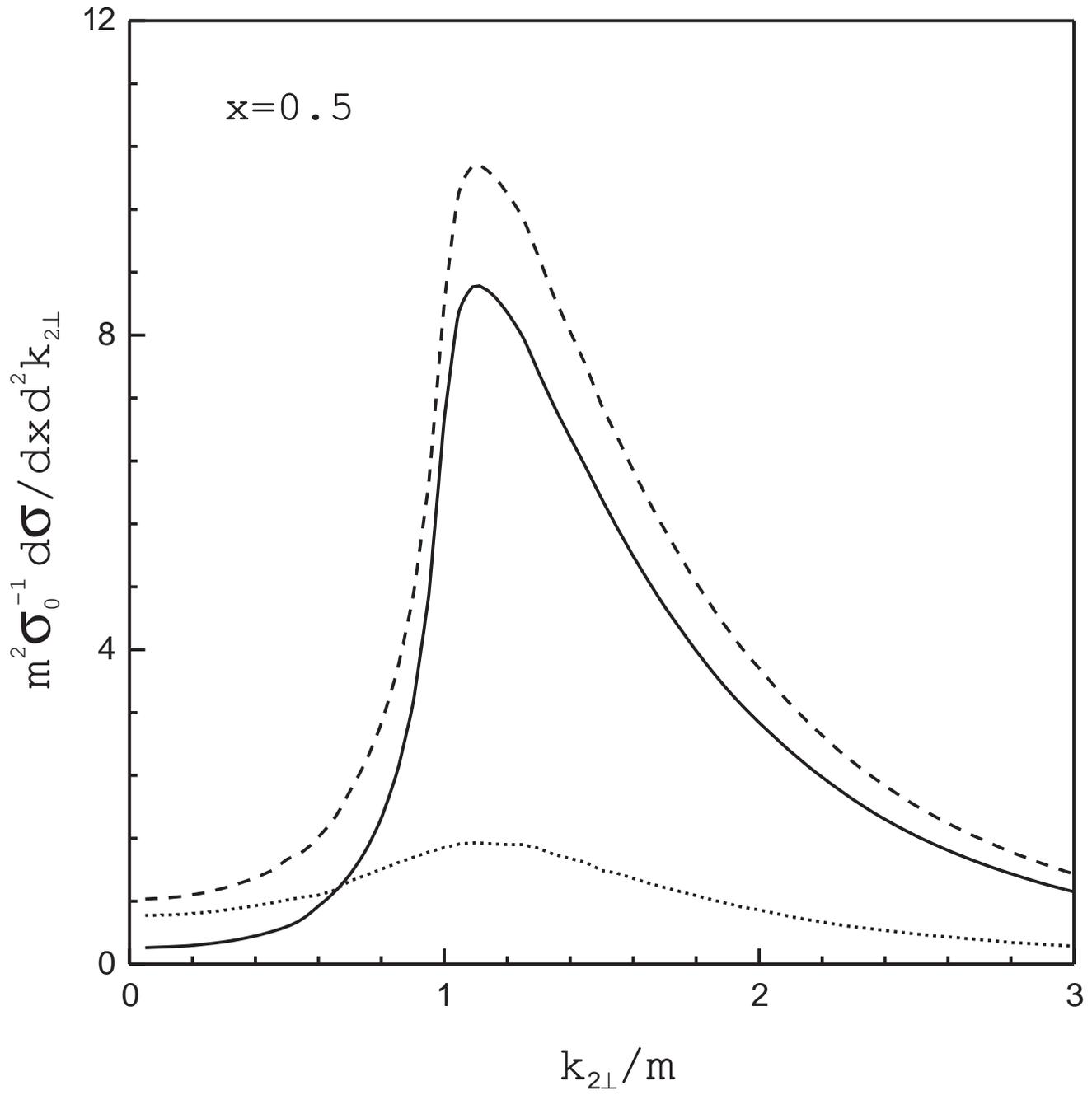
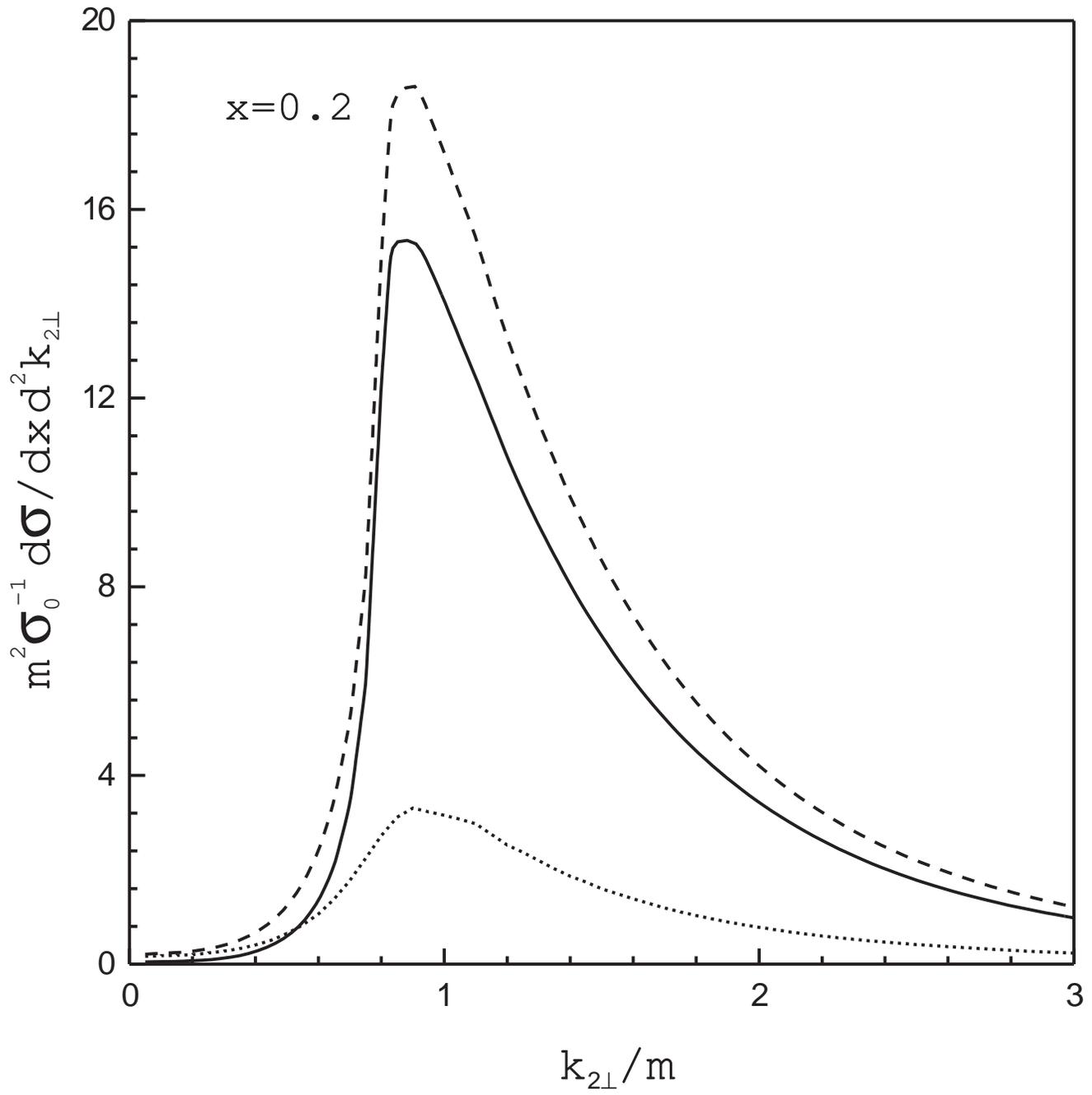
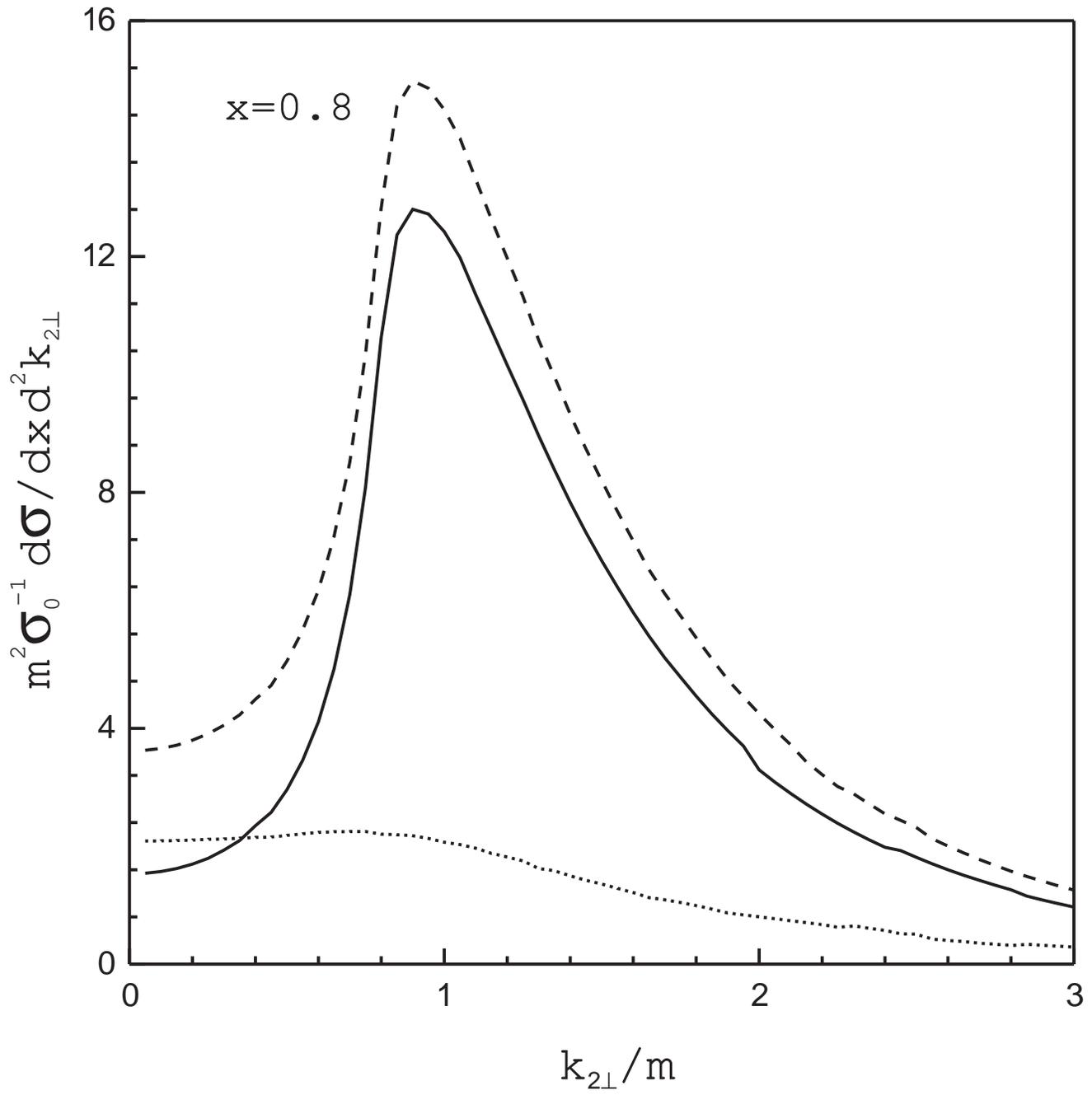
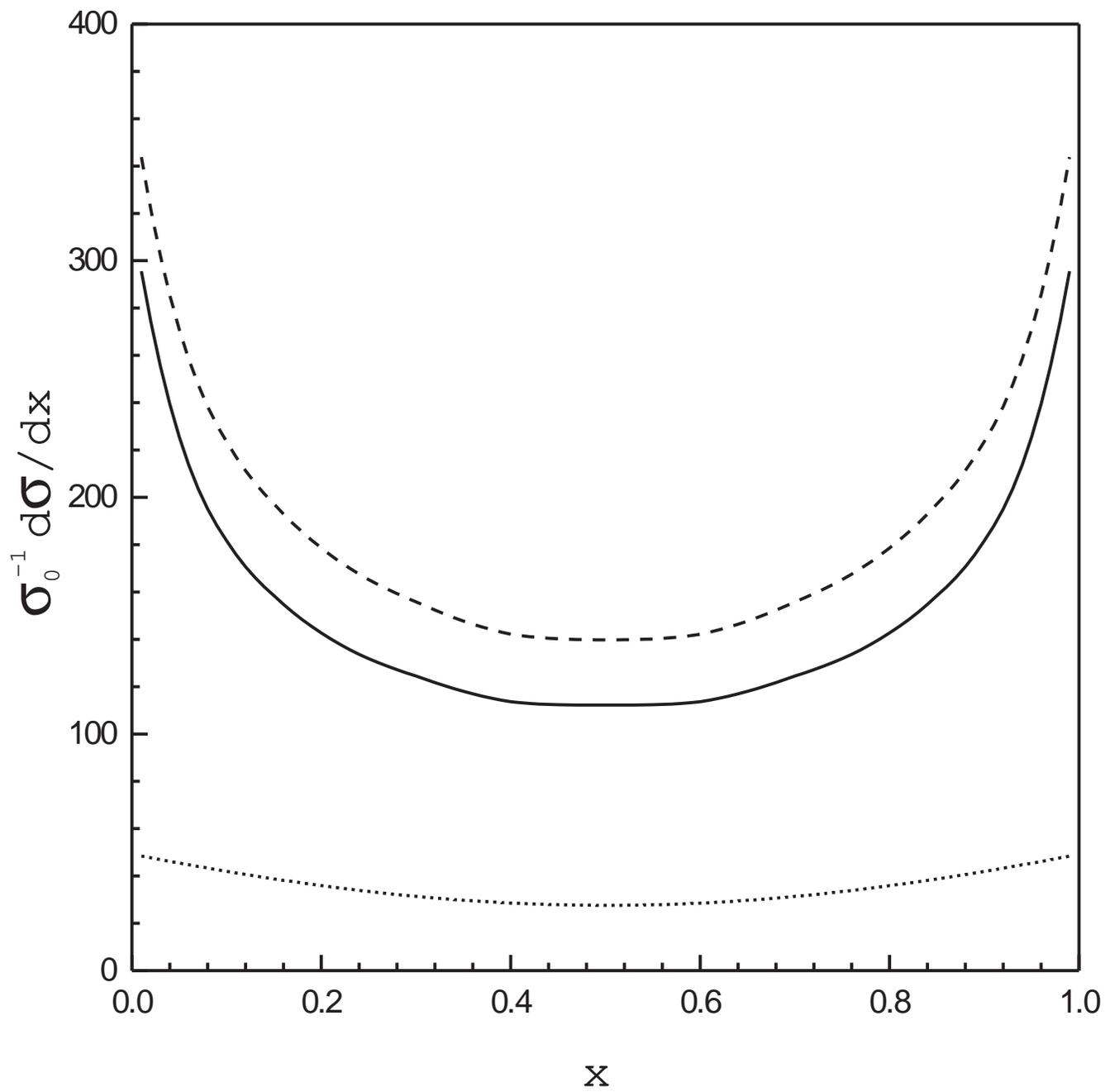
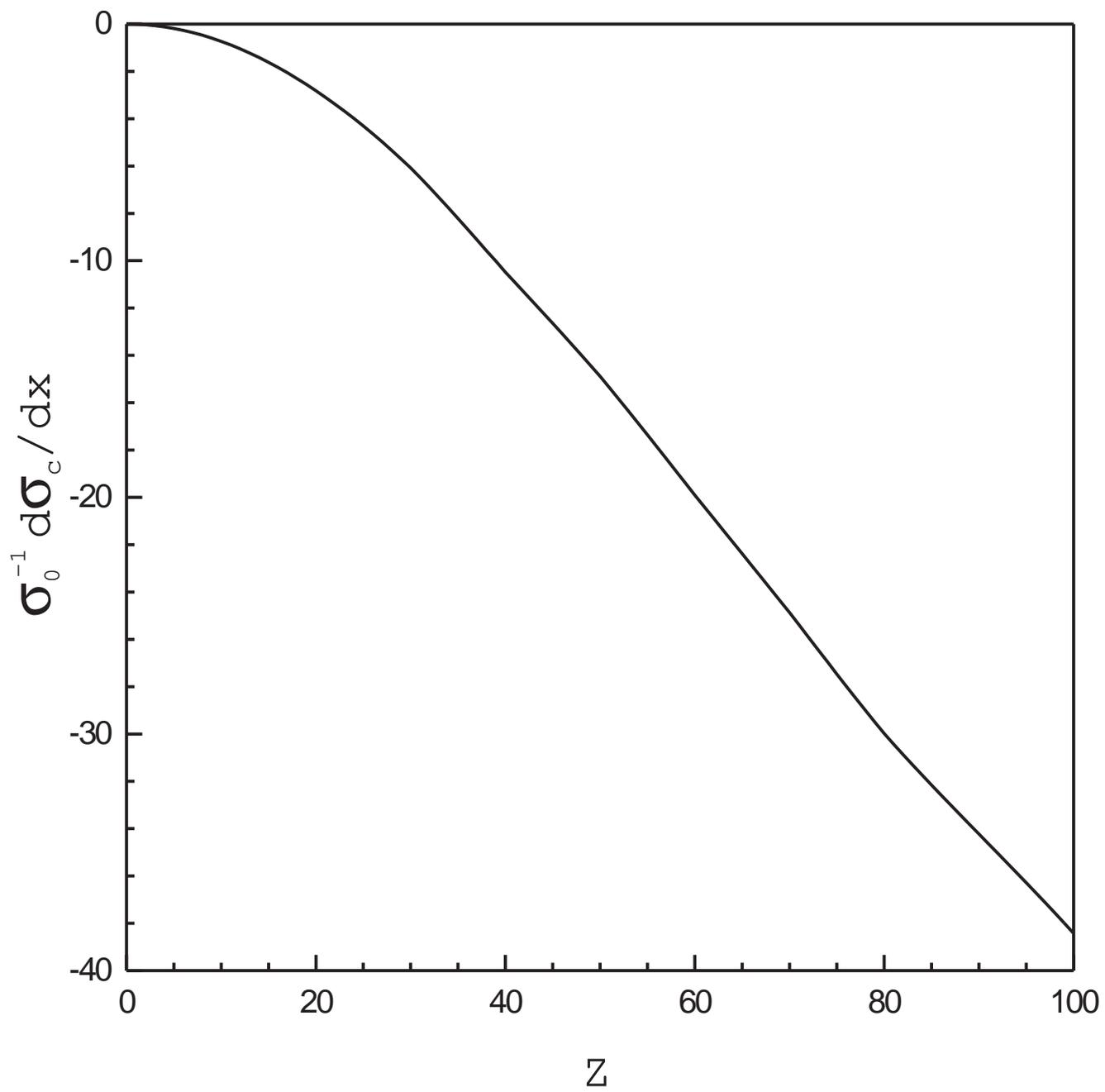
\end{document}